\documentclass{aa}  
\usepackage{graphicx}
\usepackage{txfonts}
\begin{document}
   \title{Accretion disks in Algols: Progenitors and evolution}

   \author{W. Van Rensbergen and J.P. De Greve}

   \offprints{W. Van Rensbergen}

   \institute{Astrophysical Institute, Vrije Universiteit Brussel, Pleinlaan 2, 1050 Brussels, Belgium\\
   \email {wvanrens@vub.ac.be}
   }

    \date{Received April 26, 2016; accepted June 7, 2016}

 
  \abstract
   {There are only a few Algols with measured accretion disk parameters. These measurements provide additional constraints for tracing the origin of individual systems, narrowing down the initial parameter space.}
   {We investigate the origin and evolution of six Algol systems with accretion disks to find the initial parameters and evolutionary constraints for them.}
   {With a modified binary evolution code, series of close binary evolution are calculated to obtain the best match for observed individual systems. }
   {Initial parameters for six Algol systems with accretion disks were determined matching both the present system parameters and the observed disk characteristics.}
   {When Roche Lobe Overflow (RLOF) starts during core hydrogen burning of the donor, the disk lifetime was found to be short. The disk luminosity is comparable to the luminosity of the gainer during a large fraction of the disk lifetime.}

   
    \authorrunning{W. Van Rensbergen et al.}  
     \titlerunning{Mass loss out of close binaries}
   \maketitle

\section{Introduction}

Peters (\cite{Peters}) defines an Algol as a semidetached binary in which the less massive star fills its Roche Lobe (RL) and is the cooler, fainter and larger, while the most massive star does not fill its RL and is still on the main sequence.

The evolution of a binary is conservative when mass and angular momentum of the system remain constant.
Eggleton (\cite{Eggleton}) introduced the word ``liberal'' when this is not the case.
However, it is difficult to calculate the amount of mass that leaves the system.
Wellstein (\cite{Wellstein}) assumes that mass is lost from a massive system as soon as the gainer spins at critical velocity. Van Rensbergen et al. (\cite{VanRensbergenetal1}) expand this criterion estimating the temperature of a high temperature accretion region (HTAR) so that the sum of rapid rotation and radiation pressure from the hot region overcomes gravitation and drives mass out of the binary. This criterion should be refined as soon as more observed values of sizes and temperatures of hot regions are available.

Orbital periods of Algols are known accurately. Somewhat different values of masses, radii and effective temperatures of the components are found in the literature (e.g. in the catalogs of Budding et al. \cite{Buddingetal} and Brancewicz $\&$ Dworak \cite{Brancewiczetal}).

The most important problems in modeling an individual system are that a binary evolutionary calculation starts with assumed values of initial mass of the donor (subscript $d$) and the gainer (subscript $g$) and the initial period of the system. This introduces an intrinsic uncertainty since the present values of the masses are not accurately known. Very little to nothing is known about possible present or past mass loss and subsequent loss of angular momentum from the system.

Algol systems with an accretion disk are a special subset, as the accretion disk puts additional constraints on modeling the evolution of the system. The characteristics of six of them are elaborated further in  section \ref{sec_Modeling}. We will use them to constrain the parameters of the progenitors and their evolution. We first discuss the changes that have been made to the evolution code since the one used in previous liberal binary calculations (Van Rensbergen et al. \cite{VanRensbergenetal1},~\cite{VanRensbergenetal2}) and the joined catalog.
      

\section{Modifications in the binary evolution code}
\label{sec_ModificationsCode}
\subsection{Tides}

\label{sec_Modifications}

%

Tidal interactions modulate the angular velocity of a binary member ($\omega$) with the angular velocity of the system ($\omega_{orb}$). Following Wellstein  (\cite{Wellstein}) we have

\begin{equation}
{1 \over {\omega-\omega_{orb}}} ~ {d \omega_{orb} \over  d t}= - {1\over {\tau_{sync}  f_{sync}}}= -{1 \over t_{sync}}
\label{fsyncsync} 
\end{equation}

where $t_{sync}$ is the synchronization time and $\tau_{sync}$ is a characteristic time associated with the theory of Darwin (\cite{Darwin})

\begin{equation}
\tau_{sync}~(yr)={q^{-2}} \left({a \over R}\right) ^{6}
\label{tausync} 
\end{equation}

The star that is synchronized is in the denominator of the expression for the mass ratio q. Wellstein (\cite{Wellstein}) uses  $f_{sync}$=1 for weak tides and  $f_{sync}$= 0.1 for strong tides in his scenario for liberal evolution of massive close binaries. Synchronization is strong on a star with a radius ($R$) comparable to the size of the semimajor axis of the binary ($a$).

If stars rotate asynchronously with the orbit, their spin changes by the tides, changing their angular momentum with an amount

\begin{equation}
\Delta J_{spin}={I} ~ {(\omega_{orb}-\omega)} ~ {\lbrack 1 - e^{{-\Delta t  \over  {t_{sync}}}} \rbrack}
\label{downspin}
 \end{equation}  

To avoid the artificial quantity  $f_{sync}$ we calculate $t_{sync}$ with a physical model, discriminating between convective and radiative envelopes. Hurley et al. (\cite{Hurleyetal}) use an expression of Hut (\cite{Hut}) to calculate $t_{sync}$ for circular orbits of stars with a convective envelope

\begin{equation}
{t_{sync} = {q^{-2}} ~ \left({a \over R}\right) ^{6}} ~ {1\over 3~k_{2}} ~ t_{F}   {I \over M~R^{2}}
\label{tsync} 
\end{equation}
  
where $k_{2}$ is the apsidal motion constant; $t_{F}$ is the viscous friction time; and $M$, $R$, and $I$ are respectively the stellar mass, radius and moment of inertia. Following Zahn (\cite{Zahn2}) it is possible to write~$t_{F}={R \over \nu_{t}}$, where $\nu_{t}$ is the coefficient of eddy viscosity. It is clear that $\nu_{t}$ = 0 for a star with an envelope in radiative equilibrium, so that for these stars $t_{F}$ is infinite and tidal friction is not at work.
If a convection region is a substantial fraction of a star and if convection transports most of the energy flux, the viscous friction time (Zahn, \cite{Zahn2}) is

\begin{equation}
{t_{F}(yr)=\left({{M~R^{2}} \over {L}}\right)^{1\over 3} = 0.4311~ \left({{{M \over M_{\odot}}~\lbrack{{R \over R_{\odot}}}\rbrack^{2}} \over {L \over L_{\odot}}}\right)^{1\over 3} }
\label{teef}
\end{equation}
  
Relation (\ref{tsync}) can thus be written as:

\begin{equation}
{t_{sync}(yr)=0.4311~ {q^{-2}} ~{\left( {a\over R} \right)} ^{6} ~{1\over 3~k_{2}}  {I \over M~R^{2}}  \left({{{M \over M_{\odot}}~\lbrack{{R \over R_{\odot}}}\rbrack^{2}} \over {L \over L_{\odot}}}\right)^{1\over 3} }
\label{tsyncbis}
 \end{equation}
  
During stellar evolution $k_{2}$ can be calculated from a simplified formula given by Kopal (\cite{Kopal2}), also used by Odell (\cite{Odell}) and Kopal (\cite{Kopal1}), using an integral from the center to the edge of the gainer.

\begin{equation}
{k_{2} = {16~\pi \over 5~M~R^{2}}~{\int^{R_{g}} _{0} {\rho (r)~r^{2}~dr}}}
\label{K2} 
\end{equation}
  
Hilditch (\cite{Hilditch}) gives an expression of Zahn (\cite{Zahn1}) to calculate the tidal action on a star with a radiative envelope and a convective core. Owing to the convective core the tidal torque mechanism is at work resulting from the oscillation of the convective core. The corresponding synchronization time is

\begin{eqnarray}
t_{sync}(yr)&=&3.1816 ~ 10^{-6} ~ \left({{R \over R_{\odot}}  \over {M \over M_{\odot}}}\right)^{1\over 2} ~ {I \over M~R^{2}}\nonumber \\
&\times& {1 \over E_{2}} ~ {q^{-2}} ~  {\left({a \over R}\right)^{8.5}}  {1 \over {(1+q)^{5 \over 6}}}
\label{tsyncrad}
\end{eqnarray}

The strength of the tide depends critically upon the size of the convective core. The tidal-torque constant $E_{2}$ is zero for stars with mass below 1.25 $M_{\odot}$ so that this mechanism has no effect on a star with a convective atmosphere.
 
\vspace{0.2cm}

However, the radius of the convective core is larger than given by the Schwarzschild (\cite{Schwarzschild}) criterion. Using the mixing length $\ell$ of the path after which a convective cell dissolves and the pressure scale height $H_{P}$, the overshooting parameter $\alpha$ is defined as $\alpha$ = $\ell$/$H_{P}$.

Our choice of values of $E_{2}$ calculated by Claret (\cite{Claret}) with overshooting of the convective core characterized by $\alpha$ = 0.2 is justified for the sample of binaries in this paper. We used the spectroscopic HRD proposed by Langer $\&$ Kudritzki (\cite{LangerKudritzki}). The location in this diagram of the ``red points'' (coolest points during main sequence evolution) of massive stars was taken from Castro et al. (\cite{Castroetal}) and MacDonald et al. (\cite{McDonaldetal}) for lower mass stars. We calculated the main sequence evolution of single stars between 2 and 30$M_{\odot}$ using different values of the parameter $\alpha$. We found that the choice $\alpha$  = 0.2 is justified for stars with masses below 12.5$M_{\odot}$, growing gradually up to $\alpha$ = 0.4 at 30$M_{\odot}$.

Expressions (\ref{tsyncbis}) and (\ref{tsyncrad}) refine and quantify the original theory of Darwin (\cite{Darwin}).
Belczynski et al. (\cite{Belczynskietal}) apply radiative damping to stars with radiative envelopes: Main sequence stars with mass above $M_{MS,conv}=1.15M_{\odot}$, CHeB stars above $7M_{\odot}$, massive evolved He stars, and He MS stars. For all other stars convective damping is applied. We assume that during RLOF the atmospheres of both stars are sufficiently disturbed so they can be considered to be in convective equilibrium.

\vspace{0.2cm}
Tidal action produced by meridional circulation (Tassoul \cite{Tassoul}) can be added to the Darwin theory in order to realize synchronization of binaries over a wider range than performed by the Darwin interaction alone. The synchronization time in the Tassoul theory is given by
\vspace{0.2cm}

\begin{eqnarray}
t_{sync}(yr) &=&  5.396 ~ {10^{2+\sigma-{N \over 4}} ~ {1+q \over q}} ~ {\left({{L_{\odot}}  \over L}\right)^{1 \over 4}} ~ {\left({{M\over M_{\odot}} }\right)^{3 \over 4}}\nonumber \\
&\times&{\left({{R_{\odot}}  \over R}\right)^{3}}   {\lbrack P(days) \rbrack}^{11\over 4}
\label{tsynctassoul}
\end{eqnarray}

In the same way as $t_{F}$ is the main uncertainty of the Hut theory in Eq. (\ref{tsync}), the values of N and $\sigma$ are the main uncertainties of the Tassoul theory in Eq. (\ref{tsynctassoul}).
The number N is given as a function of the coefficients of turbulent and radiative viscosity: $10^{N}= {\nu_{rad}+{\nu_{turb}} \over{ \nu_{rad}}}$. In the absence of turbulence we have for a radiative envelope N = 0. We assume with Tassoul $\&$ Tassoul (\cite{TassoulTassoul}) that N takes the rather large value N = 10 for a convective envelope,  because turbulent viscosity is always much larger than radiative viscosity if convection is at work. The always uncertain value of $\sigma$ will be calculated within these assumptions so that relation (\ref{tsynctassoul}) reproduces observations best.

Using Kepler's 3rd law $t_{sync}$ increases in relation (\ref{tsynctassoul}) with increasing separations as $\left({a\over R}\right)^{4.125}$ which is much slower than $\left({a\over R}\right)^{6}$ in relation (\ref{tsyncbis}) and $\left({a\over R}\right)^{8.5}$ in relation (\ref{tsyncrad}). 

\vspace{0.2cm}

Claret et al. (\cite{Claretetal}) propose ${\sigma}$=1.6, leading to values of the exponent respectively 3.6 (radiative) and 1.1 (convective) in relation (\ref{tsynctassoul}). We propose calculating ${\sigma}$ so as to keep the components of binaries synchronized up to orbital periods of 10 days as observed by Matthews  \& Mathieu (\cite{MatthewsMathieu}) for main sequence A stars. We calculated an extended set of binary main sequence evolutions for stars between 1 and 30 $M_{\odot}$ with lower mass companions. Initial periods were between 5 and 15 days, so that no RLOF occurred during the main sequence life of the most massive component. The binaries exerted tidal interaction on one another, starting with the Darwin tide (relation (\ref{tsyncbis}) for stars with a convective envelope and relation (\ref{tsyncrad}) for stars with a radiative envelope). The Darwin mechanism did not keep the observed fraction of binaries synchronized. We added the Tassoul tide (\ref{tsynctassoul}) using different values of ${\sigma}$ in order to achieve synchronization within reasonable limits.

\begin{figure*}[!ht]
\centering
\includegraphics[width=9.6cm]{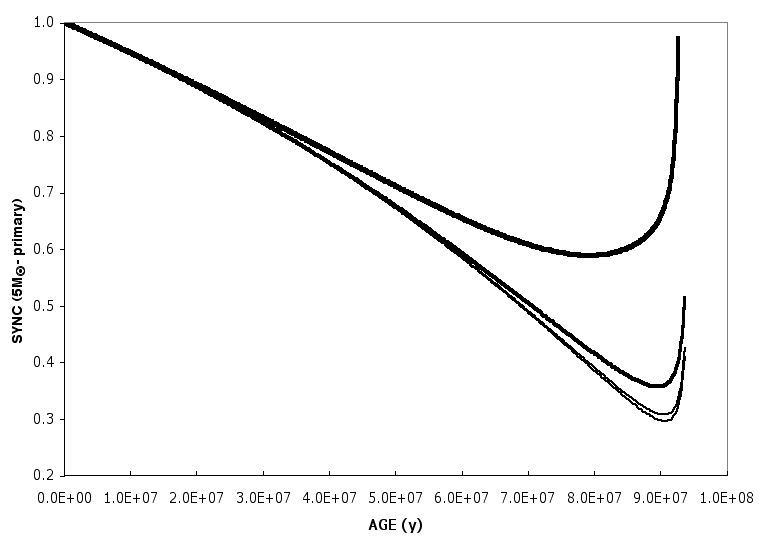}
\caption{Evolution of SYNC =  ${\omega_{spin} \over \omega_{orb}}$, during the main sequence life of the $5 M_{\odot}$ primary in a $5 M_{\odot} + 2 M_{\odot}$ binary with a 10-day orbital period. From bottom to top: without tide, Darwin tide contributing almost nothing to the synchronization, weak Tassoul tide added, strong Tassoul tide added. The $5 M_{\odot}$ gainer needs the strong Tassoul tide to remain sufficiently synchronized. }
\label{fig_fig1}
\end{figure*}

Figure \ref{fig_fig1} shows the evolution of the factor SYNC =  ${\omega_{spin} \over \omega_{orb}}$ for the primary in a binary pair $5 M_{\odot} + 2 M_{\odot}$. The system starts synchronized and the evolution is followed from ZAMS to the end of hydrogen core burning for the  primary. During this time the less massive component remains synchronized, even if no tides are at work. Without tide, the  $5 M_{\odot}$ star is soon asynchronized. Accurate inspection of Figure \ref{fig_fig1} shows that the Darwin tide contributes almost nothing to the synchronization of the $5 M_{\odot}$ primary. The contribution of the weak Tassoul tide (${\sigma}=4.5$) is not of very much help. The Darwin tide enforced by the strong Tassoul tide (${\sigma}=3.5$) restores synchronization most of the time for this system with an orbital period of 10 days. Therefore, the value  $2+ {\sigma} - {N \over 4} = 5.5$ will be used in relation (\ref{tsynctassoul}) for the tidal tuning of a star with a radiative atmosphere.

In order to find a value for $\sigma$ for a star with a convective atmosphere we followed the evolution of a binary pair $1.15 M_{\odot} + 1 M_{\odot}$. We find that without tide the $1.15 M_{\odot}$ star is soon asynchronized. Convective Darwin tides restore synchronization for a 10 d orbital period without help of the Tassoul tide. However, for a 15 d orbital period, the Darwin tide restores synchronization less well. The Tassoul tide in its weak form (${\sigma}=5$) restores synchronization completely. Therefore, we use the value  $2+ {\sigma} - {N \over 4} = 4.5$ in relation (\ref{tsynctassoul}) for the tidal tuning of stars with a convective atmosphere.

The Tassoul and Darwin mechanisms act simultaneously, so that the resulting synchronization times are given by
 					
\begin{equation}
{{1 \over t_{sync}} = {1 \over t_{sync,Darwin}} + {1 \over t_{sync,Tassoul} }}
\label{tcomb}
 \end{equation} 

\subsection{Conservation of angular momentum}

The total angular momentum $J_{\Sigma}$ of a binary is the sum of the orbital angular momentum $J_{orb}$  and the spin angular momenta of gainer $J_{g}$  and donor $J_{d}$ . Tides continuously exchange amounts of  $\Delta J_{d}$ , $\Delta J_{g}$ ,$\Delta J_{orb}$ . In the conservative case\\

$\Delta J_{\Sigma} = \Delta J_{d}+ \Delta J_{g}+ \Delta J_{orb} = 0$.\\

However, the system loses angular momentum through stellar wind ($\Delta J_{wind} < 0$), and during its liberal era also through mass loss out of the system ($\Delta J_{out} <0 $). Conservation of angular momentum during evolution is then imposed by
 
\begin{equation}
{{\Delta}J_{d} + {\Delta}J_{g} + {\Delta}J_{orb} - {\Delta}J_{wind} - {\Delta}J_{out} = 0}
\label{DeltaJlib}
 \end{equation} 
 
In this paper, we calculated $\Delta J_{wind}$ using Vink et al. (\cite{Vinketal}) for stars hotter than 12500 K and De Jager et al. (\cite{Dejageretal}) for cooler stars. In the case of liberal evolution the quantity  $\Delta J_{out}$ was calculated as in previous papers (Van Rensbergen et al. \cite{VanRensbergenetal1}, \cite{VanRensbergenetal2}; see also Siess et al. \cite{Siessetal}) assuming that mass lost from the system takes only the specific orbital angular momentum of the gainer.

\section{Obsevations}
\label{sec_Obs}

The observed data for which we calculated the most plausible progenitors are given in Tables \ref{tab_tab1} (system parameters) and \ref{tab_tab2} (disk parameters). 
Table \ref{tab_tab1} gives the observed system parameters to be met by the evolution of our progenitors. Masses and radii of $\beta$~Lyr are from Zhao et al. (\cite{Zhaoetal}), the effective temperatures and mass transfer rates from Harmanec (\cite{Harmanec}). Masses, radii and effective temperatures for AU Mon are from Desmet et al. (\cite{Desmetetal}) and Atwood-Stone et al. (\cite{Atwoodetal}). We determined a very uncertain mass transfer rate applying the (O-C) method on the data found in the catalog of Kreiner et al. (\cite{Kreineretal}). Masses, radii and effective temperatures for V356 Sgr are from Dominis et al. (\cite{Dominisetal}). The mass transfer rate is from Ziolkowski (\cite{Ziolkowski}). Masses, radii and effective temperatures for TT Hya are from Miller et al. (\cite{Milleretal}). The mass transfer rate was determined from the data found in the catalog of Kreiner et al. (\cite{Kreineretal}). Masses, radii and effective temperatures for RY Per are from Peters $\&$ Polidan (\cite{PetersPolidan}). No trustworthy determination of the mass transfer rate was found in this case. Masses and radii for SW Cyg are from Richards $\&$ Albright (\cite{RichardsAlbright}), the effective temperatures from Budding et al. (\cite{Buddingetal}) whereas the mass transfer rate is from Qian et al. (\cite{Qianetal}).

\begin{table*}
\begin{center}
\begin{tabular}{ccccccccc} \hline
System & P& $M_{d}$ & $M_{g}$ & $R_{d}$ & $R_{g}$ & Log $T_{eff,d}$  & Log $T_{eff,g}$ & ${dM/dt}$  \\ \hline
Units & days & $M_{\odot}$ &$M_{\odot}$& $R_{\odot}$ & $R_{\odot}$ &   &  & $M_{\odot}/ yr$ \\ \hline
\hline
${\beta~Lyr}$ & 12.91378 & 2.88 & 12.97 & 14.7 & 6.1 & 4.114 & 4.447 & 2.03E-5  \\
${AU~Mon}$ & 11.11304 & 1.2 & 7.0 & 10.0 & 5.6 & 3.76 & 4.23 & 2.59E-6  \\
${V356~Sgr}$ & 8.89611 & 2.8 & 10.4 & 11.7 & 5.2 & 3.954 & 4.362 & 1.90E-6  \\
${TT~Hya}$ & 6.95343 & 0.63 & 2.77 & 5.98 & 1.99 & 3.68 & 3.99 & 6.85E-8  \\
${RY~Per}$ & 6.68356 & 1.6 & 6.25 & 8.1 & 4.06 & 3.802 & 4.259 &   \\
${SW~Cyg}$ & 4.57302 & 0.5 & 2.5 & 4.3 & 2.6 & 3.69 & 3.956 & 2.11E-7  \\
\hline
\end{tabular}
\caption{Adopted observed values of masses, radii, effective temperatures and mass loss rates of six Algols with accretion disks.The references are given in section \ref{sec_Obs}.}
\label{tab_tab1}
\end{center}
\end{table*}

Table \ref{tab_tab2}  compares observed luminosities, temperatures and sizes of the six disks with the same quantities obtained in section \ref{sec_Modelingdisk} from the best models. Disk luminosities for SW Cyg and TT Hya are taken from Albright $\&$ Richards (\cite{AlbrightRichards}). The outer edge temperature of AU Mon is from Djurasevic et al. (\cite{Djurasevicetal}). Other characteristic disk temperatures are from Miller et al. (\cite{Milleretal}) for TT Hya, from Harmanec (\cite{Harmanec}) and Ak et al. (\cite{Aketal}) for $\beta$~Lyr and from Wilson \& Caldwell (\cite{WilsonCaldwell}) for V356 Sgr. The sizes are from Atwood-Stone et al. (\cite{Atwoodetal}) for AU Mon, Albright $\&$ Richards (\cite{AlbrightRichards}) for SW Cyg, Peters (\cite{Peters1}) for TT Hya, Sudar et al. (\cite{Sudaretal}) for RY Per, Mennickent $\&$ Djurasevic (\cite{MennickentDjurasevic}) for $\beta$~Lyr and from Wilson $\&$ Caldwell (\cite{WilsonCaldwell}) for V356 Sgr. The size of the disk in Table \ref{tab_tab2} is the disk radius divided by the Roche radius of the gainer.  Since accretion disks in Algols are seen almost edge on, the model temperatures that were calculated at the edge show a fair agreement with the observations in table \ref{tab_tab2}.

\begin{table*}
\begin{center}
\begin{tabular}{cccccccc} \hline
System & Obs.$L_{disk}\over L_{g}$ & Calc.$L_{disk}\over L_{g}$ & Obs.Temp & Calc.$T_{edge}$ & Calc.$T_{eff,disk}$ & Obs.Size  & Calc.Size  \\ \hline
${\beta~Lyr}$ & & 0.713 & \lbrack 8500-9000 \rbrack & 8617 & 15233 & 0.91 & 0.90 \\
${AU~Mon}$ & & 0.666 & 5190 & 6546 & 11510 & 1.00 & 0.90  \\
${V356~Sgr}$ & & 0.293 & \lbrack 6000-7000 \rbrack  & 11040 & 15610 & 0.65 & 0.50 \\
${TT~Hya}$ & 0.414 & 0.903 & 7000 & 3110 & 5858 & 0.95 & 0.86  \\
${RY~Per}$ & & 0.616 & & 7969 & 12456 & 0.85 & 0.80   \\
${SW~Cyg}$ & 0.096 & 0.406 & & 4093 & 6364 & 0.95 & 0.60  \\
\hline
\end{tabular}
\caption{Observed and calculated disk characteristics. The size is the radius of the disk divided by the Roche radius of the gainer. Only for AU Mon the observed temperature is at the edge. The references are given in section \ref{sec_Obs}}
\label{tab_tab2}
\end{center}
\end{table*}

\section{Modeling six Algols with accretion disks}
\label{sec_Modeling}

\subsection{Modeling the stars}
\label{sec_Modelingstars}

Using our binary evolutionary code with the modifications discussed in section \ref{sec_ModificationsCode} we determine in this subsection the progenitors of six binaries with an accretion disk around the gainer. Four of them turn out to evolve conservatively (AU Mon, SW Cyg, TT Hya and RY Per), whereas $\beta$~Lyr and V356 Sgr have liberal eras during their evolution. In the conservative case the initial orbital period of a system follows directly from the present situation (subscript 1) for any combination of the initial masses (subscript 0).

\begin{equation}
{{{P_{1}} \over {P_{0}}} = \left({{M_{d,0} ~M_{g,0}} \over {M_{d,1}~ M_{g,1}}}\right)^{3} }
\label{Pcons}
\end{equation}
 
This relation disregards small deviations that are caused by the stellar wind of both components and the action of the tides. {Mennickent (\cite{Mennickent}) proposes a (4 $M_{\odot}$ + 3.6 $M_{\odot}$, P=3d}) progenitor for AU Mon with a 1.533 $M_{\odot}$ donor and a 6.067 $M_{\odot}$ gainer at present. Since we model AU Mon towards a  1.2 $M_{\odot}$ donor and a 7 $M_{\odot}$ gainer our progenitors with initial mass ratio ({4$/$3.6}) need to have an initial orbital period a little below 1.4 d, determined from expression (\ref{Pcons}). With this small initial orbital period the stars in the binary merge soon after birth.

All the gainers in Table \ref{tab_tab1} are main sequence stars whereas donors are evolved beyond the main sequence. 

Initial orbital periods of conservative cases were calculated with expression (\ref{Pcons}).

\vspace{0.2cm}

Loss of mass in the liberal cases is triggered by the combined action of rapid rotation and radiation pressure from the HTAR region where the transformation of matter energy into radiation is crucial. The radiative efficiency is defined through the relation: $L_{add}$~=~$\eta$  ${\dot M}$ $c^{2}$. For comparison we give some numbers: $\eta$ $\approx$ 0.007 for nuclear fusion, $\eta$ $\approx$ $10^{-4}$ for an accreting white dwarf, $\eta$ $\approx$ $10^{-2}$ for an accreting neutron star. The value of $\eta$ for a hot spot on a main sequence gainer in the case of direct impact was calculated by Van Rensbergen et al. (\cite{VanRensbergenetal2}) and found to be significantly smaller than in the case of a mass accreting white dwarf. In the case of a gainer surrounded by a disk, the hot spot could be located at the edge of the disk. Moreover, Bisikalo (\cite{Bisikalo}) observed and reproduced hot lines using 2D and 3D gas dynamical simulations for disk systems. Using the same criterion as used by Van Rensbergen et al. (\cite{VanRensbergenetal2}) we found that four out of six systems treated in this paper evolved conservatively. The two liberal systems (V356 Sgr and $\beta$~Lyr) were observed to be ejecting mass into space.

\vspace{0.2cm}

Initial periods for  liberal cases were calculated with Eq. (\ref{DeltaJlib}) assuming a loss of specific angular momentum $\Delta J_{out}$ equal to the orbital angular momentum of the gainer. The mass loss out of the system was found to be vertical on the orbital plane and has been interpreted as bipolar jets for V356 Sgr by Peters \& Polidan (\cite{PetersPolidan}) and for $\beta$~Lyr  by Ak et al. (\cite{Aketal}) and Harmanec (\cite{Harmanec}). Mass loss through the second Lagrangian point is unlikely because this point is far away from the outer edge of the accretion disk, and the large amount of angular momentum lost narrows the orbit so that the formation of an accretion disk around the gainer is inhibited.

Liberal cases were calculated with various amounts of mass lost by the system. However, assuming a mass of [1.5-2] $M_{\odot}$ lost by $\beta$~Lyr and V356 Sgr gives the best fit with the observed data.

The progenitor systems are characterized by $(P_{0}, M_{d,0}, M_{g,0})$. Evolutionary sequences were calculated for various combinations of $(M_{d,0}, M_{g,0})$ and evaluated at P = ${\mathrm{P_{obs}}}$. The results are shown in Table \ref{tab_tab3}.

All the binaries in Table \ref{tab_tab1}  have a mass ratio q $<$ 0.27. For those binaries that started RLOF during core hydrogen burning of the donor this means that they are at the end of their RLOF evolution. TT Hya and RY Per are not in that case.

\subsection{Modeling the disk}
\label{sec_Modelingdisk}

Our code also reproduces characteristics of the accretion disk around the gainer. The criterion of Lubow $\&$  Shu (\cite{LubowShu}) uses the concept of fractional radius $\varpi$ which is the radius divided by the semimajor axis. The fractional radius is determined at the edge of the gainer $\varpi_{g}$ and at specific distances  $\varpi_{d}$  \footnote{This subscript "d" does not have the same meaning as used for the donor.} and  $\varpi_{min}$ so that a semidetached binary changes from direct hit into accretion through a transient disk as soon as $\varpi_{g}$ =  $\varpi_{d}$ and later on through a permanent accretion disk when $\varpi_{g}$ =  $\varpi_{min}$. 
At $\varpi_{d}$ the disk radius starts from $R_{disk} = R_{g}$ until the radius of a permanent disk $R_{PD}$  is achieved at $\varpi_{min}$. For the radius of a permanent disk we assumed, as do Carroll $\&$ Ostlie (\cite{Carroll}), that $R_{PD} = 2~ r_{circ}$, a value that is a somewhat larger than 2~$\varpi_{d}$.

The radius of the transient disk is calculated with the assumption that the disk radius grows faster than linearly from $\varpi_{d}$ and $R_{g}$ towards $\varpi_{min}$ and $R_{PD}$. A slower growth yields sizes of transient disks that are smaller than observed.

\begin{figure*}[!ht]
\centering
\includegraphics[width=9.6cm]{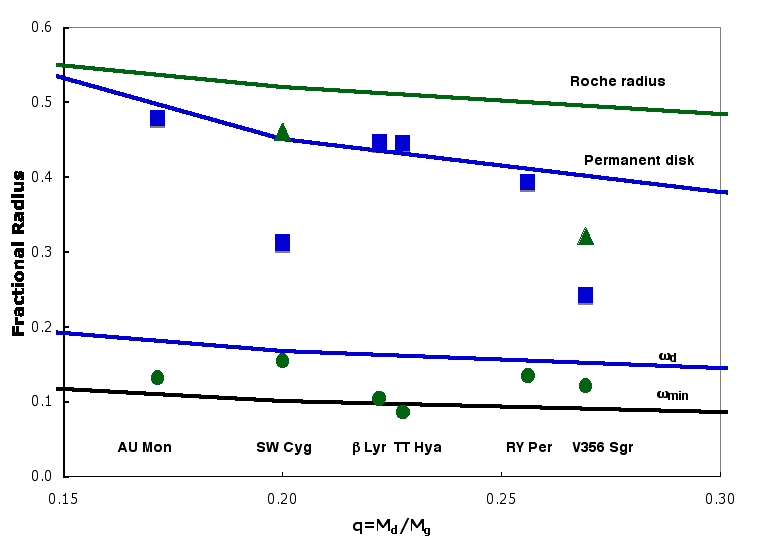}
\caption{Fractional radii (radius / semimajor axis) of six gainers and their accretion disks in the Lubow \& Shu diagram. The observed radii of gainers are dots below $\varpi_{d}$. The calculated disk radii are squares above $\varpi_{d}$. The two observed disk radii are triangles. The fractional radii of permanent disks and the position of the gainer's Roche radius are also indicated.}
\label{fig_fig2}
\end{figure*}

The Lubow \& Shu Figure $\ref{fig_fig2}$ shows the fractional radii of the six Algol gainers in this paper. All the observed fractional radii of the gainers are located below $\varpi_{d}$ as needed to develop an accretion disk. TT Hya is located below  $\varpi_{min}$. The calculated disk radii are listed in Table \ref{tab_tab2}. The gainers of four systems are surrounded by a transient accretion disk. TT Hya has a permanent disk and $\beta$~Lyr is at the transition between the transient and permanent disk. The observed size of the accretion disk in SW Cyg is much larger than the calculated one. Since the present position of SW Cyg in Figure \ref{fig_fig2} suggests that SW Cyg is far from the permanent disk phase, we conclude that the observed size overestimates reality.

The temperature in the disk is both caused by accretion and by the luminosity of the gainer. The accretion luminosity of a disk can be approximated with
 
\begin{equation}
{{L_{acc,disk}  \over {L_{\odot}}} = {{G~{M_{g}}~{\dot M_{d}}}  \over {2~{R_{g}}}}  =  1.57066 ~10^7~  {{{ \left({M_{g} \over M_{\odot}}\right)}~{\dot M_{d}} \left(in {M_{\odot} \over yr}\right)}  \over {{ \left({R_{g} \over R_{\odot}}\right)}}}  }
\label{Laccsimple}
\end{equation}

This expression can be calculated even if the gainer has no disk. Hence, it is better to calculate the accretion luminosity considering Bath $\&$ Pringle (\cite{BathPringle}), Horne (\cite{Horne}) and Rutten et al. (\cite{Ruttenetal}) who give the distribution of the accretion temperature through a disk around a white dwarf or a neutron star, using a characteristic disk temperature $T_{disk}$.
With masses and radii in solar units and mass transfer rates in ${M_{\odot}}$/y, this characteristic disk temperature created by accretion is defined by

\begin{equation}
{T_{disk}= 478074 ~  \left( {{M_{g}~ \dot M_{d}} \over {R_{g}^3}} \right)^{0.25} }
\label{Tdisk}
\end{equation}
 
The distribution of the temperature through the disk is given by  

\begin{equation}
{{T(r)}^{4} = {T_{disk}^{4}} ~{\left({{R_{g}} \over {r}}\right)^{3}} \left({1 - {\sqrt {{R_{g}  \over {r}}}}}\right) }
\label{Teer}
\end{equation}

Integrating over both sides of the disk, the amount of accretion luminosity in a ring with width $dr$ at a distance r from the gainer is

\begin{equation}
{dL = 4~ \pi ~{\sigma_{R}}~ r dr {T(r)^{4}} {erg~s^{-1}}}
\label{Lum}
\end{equation}

Integrating this quantity from the gainer's edge to the outer edge of the disk, one finds
 
\begin{equation}
{{L_{acc} \over L_{\odot}} = 9.0203 ~10^{-16} ~ {T_{disk}^{4}} ~ \left({R_{g} \over R_{\odot}}\right)^{3}~ \left({{{1 \over {3~{R_{g} \over R_{\odot}}}}}} - {{1 -{2 \over 3 }~\sqrt {R_{g} \over R_{disk}}} \over {{R_{disk}} \over {R_{\odot}}}}\right)  }
\label{Lacc}
\end{equation}
 
Restricting the luminosity of the disk to only accretion yields disk temperatures that are far below the temperatures listed in Table \ref{tab_tab2}.

\vspace{0.2cm}
In Algols hot disks surround hot gainers, whereas cooler gainers yield cooler disks. 
\vspace{0.2cm} 

Moreover, the temperature at the inner edge of the disk needs to equal the effective temperature of the gainer, a requirement that is not fulfilled by relation (\ref{Teer}). The temperature of the disk is thus also influenced by the outgoing radiation of the gainer. If the local luminosity in a disk is weakened only because of its distance from the source (i.e. there is no absorption), the temperature distribution is

\begin{equation}
{T_{rad,0} = T_{eff} \left({R_{g} \over {r}}\right)^{0.5} }
\label{Trad}
\end{equation}

Using this zero-order temperature distribution and integrating from the edge of the gainer to the outer edge of the disk, the radiation luminosity obtained is

\begin{equation}
{{L_{rad} \over L_{\odot}} = 9.0203  ~ 10^{-16}  ~ {T_{eff,g}^{4}} ~ \left({R_{g} \over R_{\odot}}\right)^{2} ~ ln~ {R_{disk} \over R_{g}} } 
\label{Lrad1} 
\end{equation}

This expression overestimates the contribution of the radiation since disk matter will prevent outgoing radiation from traveling freely. Moreover, the temperatures measured at the outer edge of the disk are far below those calculated with this assumption. Unfortunately, an evaluation of the temperature at the outer edge has only been reported in the case of AU Mon (Djurasevic et al. \cite{Djurasevicetal}). Using their numbers the following is obtained

\begin{equation}
{{{T_{rad,edge,real}} \over {T_{rad,edge,0}}} = 0.6 }
\label{Tradgauge}  
\end{equation}

Since  $T_{rad,edge,0}$ can easily be calculated with relation (\ref{Trad}) we have adopted relation (\ref{Tradgauge}) to evaluate $T_{rad,edge,real}$ for the five other systems. However, table \ref{tab_tab2} shows only a moderate agreement between observed and calculated luminosities and temperatures of the disks. In order to improve the model, more and better determined values of these quantities are needed.

\vspace{0.2cm}

For our final model, we approximate the temperature distribution in the disk adding an exponentially decreasing function starting at the gainer's surface (h=0) where h is the distance above the gainer's surface

\begin{equation}
{T_{rad}(h) =  T_{eff,g}~\left({R_{g} \over {r}}\right)^{0.5}e^{-h \over H} }
\label{Tdistr} 
\end{equation}

This temperature scale height H is determined so that expression  (\ref{Tdistr}) meets the effective temperature of the gainer at h=0 and $T_{rad,edge,real}$ at the outer edge of the disk

\begin{equation}
{H~=~{{-~\left(R_{disk} - R_{g}\right)} \over {ln~0.6}}}
\label{DefH} 
\end{equation}

In this case, the integrated luminosity of the disk is
 
\begin{equation}
{{L_{rad} \over L_{\odot}} = 9.0203  ~ 10^{-16}  ~ {T_{eff,g}^{4}} ~ \left({R_{g} \over R_{\odot}}\right)^{2} ~ e^{{4~R{g}}\over {H}} {\int^{R_{disk}} _{R_{g}}~}{e^{-4~r\over H}\over r}dr}
\label{Lrad2} 
\end{equation}

The integral in relation (\ref{Lrad2}) is the difference of the values of the exponential integral Ei(${- 4}r\over {H}$) at ($r~=~R_{disk}$) and ($r~=~R_{g}$). The radiative luminosity is added to the accretion luminosity  (\ref{Lacc}) in order to obtain the real luminosity. The effective temperature of the disk, as given in Table \ref{tab_tab2}, is calculated with

\begin{equation}
{T_{eff,disk}=5770~{{{\left(L_{disk} \over L_{\odot}\right)}^{1\over 4}}~{{\left(R_{\odot} \over R_{disk}\right)}^{1\over 2}}}~\sqrt[4]{2}}
\label{Leff_disk}
\end{equation}

\section{Evolution of the individual systems}
\label{sec_Evolution}

Table \ref{tab_tab3} shows the characteristics of the initial binaries evolving best into the present characteristics given in Table \ref{tab_tab1}. The evolutionary sequences possibly fitting the present observations were evaluated at the moment that the present orbital period was attained. The best set of initial parameters is always surrounded by a few different initial systems that evolve into current comparable results. However, progenitors with somewhat smaller initial mass of the donor do not develop the observed accretion disk, whereas future donor stars with a somewhat larger initial mass merge with the gainer before an accretion disk can be formed. In general we find that the masses of the initial system are defined within  $0.2 M_{\odot}$. We find that the initial periods are in the range [1.6-2.7] d, whereas the present orbital periods are in the range [4.5-13] d. The initial mass ratios are in the range q = $M_{d} \over M_{g}$ $\in$ [1.48-2.75] whereas the present mass ratios are in the range  [0.17-0.27]. The present state of the binaries is thus far behind the orbital period minimum when both components of the binary have equal masses.

Table \ref{tab_tab4} illustrates the disk behavior and the life of the binary as an Algol. Table \ref{tab_tab4} mentions the duration of RLOF, and the percent of time that the disk appeared during RLOF. The percent of the total time of disk appearance that is shown by the present disk is also included. The binary system is always an Algol when the gainer is surrounded by an accretion disk. Four systems (AU Mon, $\beta$~Lyr, V356 Sgr and SW Cyg) underwent previously RLOF during core hydrogen burning of the donor. Their disks live for a short fraction of the RLOF era. The other two systems (TY Hya and RY Per) start RLOF when hydrogen is exhausted in the core of the donor star. Their disks live for a large fraction of the following case B of RLOF. For all six cases, the present binary has a gainer still on the main sequence and a donor with a hydrogen exhausted core. The binary is an Algol during the largest part of the RLOF era. Inspection of Table \ref{tab_tab4} also shows that the four systems that lived their first RLOF during hydrogen core burning of the donor are currently (during their second RLOF) at the end of their evolution ($\beta$~Lyr, AU Mon, V 356 Sgr and SW Cyg). The two systems that undergo their first and only RLOF during hydrogen shell burning of the donor are in the middle of their RLOF life (TT Hya and RY Per).

\begin{table*}
\begin{center}
\begin{tabular}{cccccccccc} \hline
System & Init. mass & $P_{init}$ & $M_{d}$ & $M_{g}$ &$R_{d}$ & $R_{g}$ & Log $T_{eff,d}$  & $T_{eff,g}$ & ${dM/dt}$  \\ \hline
Units &  $M_{\odot}$ & d & $M_{\odot}$& $M_{\odot}$& $R_{\odot}$ & $R_{\odot}$ &   &  & $M_{\odot}/yr$ \\ \hline
1 &  2 & 3 & 4 & 5 & 6 & 7 &  8 & 9 & 10 \\ \hline
\hline
${\beta~Lyr}$ & 10.35+7 & 2.3625 & 2.56 & 12.96 & 14.97 & 6.62 & 4.13 & 4.44 & 1.5E-5  \\
${AU~Mon}$ & 5.7+2.5 & 2.2762 & 1.12 & 7.00 & 9.96 & 5.58 & 3.93 & 4.31 & 5.6E-7  \\
${V356~Sgr}$ & 8.95+5.75 & 2.6051 & 2.65 & 10.08 & 11.76 & 5.54 & 4.13 & 4.39 & 6.2E-6  \\
${TT~Hya}$ & 2.3+1.1 & 2.2819 & 0.59 & 2.81 & 5.84 & 1.95 & 3.67 & 4.06 & 3.2E-8  \\
${RY~Per}$ & 5.25+2.6 & 2.6987 & 1.53 & 6.31 & 7.98 & 3.51 & 3.82 & 4.34 & 1.4E-5   \\
${SW~Cyg}$ & 2.2+0.8 & 1.6383 & 0.50 & 2.50 & 4.17 & 2.26 & 3.69 & 4.01 & 4.4E-9  \\
\hline
\end{tabular}
\caption{Progenitors that fit the observations of Tables \ref{tab_tab1} and \ref{tab_tab2} most closely. Initial values are at the left (columns 2 and 3). Present values generated by the model are in the last seven columns 4-10.}
\label{tab_tab3}
\end{center}
\end{table*}

\begin{table*}
\begin{center}
\begin{tabular}{cccccccc} \hline
System & Pre Algol & Algol & Algol & Algol & RLOF & Disk Lifetime $\%$ & Present Disk $\%$ \\ \hline
 & Direct hit & Direct hit & Transient Disk & Permanent Disk & Total &  RLOF & Disk appearance \\ \hline
${\beta~Lyr}$ & 1.680E5 & 9.425E6 & 4.200E4 & 8.400E4 & 9.719E6 & 1.296 & 0.2698  \\
${AU~Mon}$ & 4.161E5 & 2.838E7 & 8.200E5 & 6.840E5 & 3.030E7 & 4.963 & 0.4325  \\
${V356~Sgr}$ & 1.870E5 & 8.347E6 & 1.000E5 & 1.400E5 & 8.774E6 & 2.735 & 0.1530  \\
${TT~Hya}$ & 7.900E5 & 1.553E6 & 2.045E6 & 1.400E7 & 1.839E7 & 87.259 & 0.2113  \\
${RY~Per}$ & 9.020E4 & 2.900E4 & 2.400E4 & 6.320E5 & 7.752E5 & 84.623 & 0.0115   \\
${SW~Cyg}$ & 4.222E6 & 2.739E8 & 3.400E7 & 2.400E7 & 3.361E8 & 17.256 & 0.3345  \\
\hline
\end{tabular}
\caption{Duration of different eras during RLOF. Pre Algol, Algol and disk eras of the six progenitors in this paper (in years). Percent of RLOF that the disk is present. Fractional age of present disk.}
\label{tab_tab4}
\end{center}
\end{table*}

For each of the systems we calculated the evolution of the luminosity of the disk as a function of time. The accretion luminosity is calculated with relations (\ref{Laccsimple}) and (\ref{Lacc}). The evolution with time of the accretion luminosity reflects the evolution of the mass loss rate. 

\begin{figure*}[!ht]
\centering
\includegraphics[width=9.6cm]{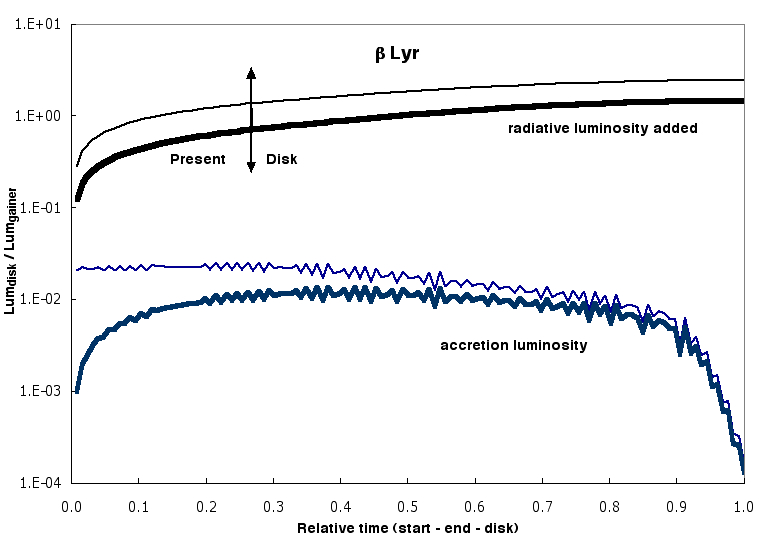}
\caption{Evolution of the luminosity of the short-lived disk of $\beta$ ~Lyr divided by the gainer's luminosity as a function of relative disk time. From bottom to top: accretion luminosities (\ref{Laccsimple}), (\ref{Lacc}), radiative luminosities (\ref{Lrad2}) and (\ref{Lrad1}) added.}
\label{fig_fig3}
\end{figure*}

\begin{figure*}[!ht]
\centering
\includegraphics[width=9.6cm]{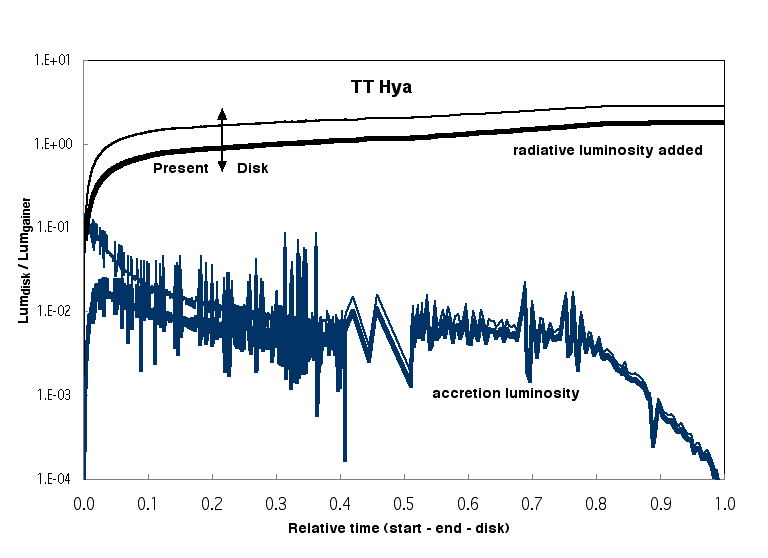}
\caption{Evolution of the luminosity of the long-lived disk of TT Hya divided by the gainer's luminosity as a function of relative disk time. From bottom to top: accretion luminosities (\ref{Laccsimple}), (\ref{Lacc}), radiative luminosities (\ref{Lrad2}) and (\ref{Lrad1}) added.}
\label{fig_fig4}
\end{figure*}

The radiation luminosity that contributes most to the luminosity of the disk is triggered by the radiation from the gainer that is absorbed in the disk and reradiated. Two approaches can be considered: the optically thin approach given by relation (\ref{Lrad1}) and the optically thick approach given by relation (\ref{Lrad2}). The resulting effect of the two approaches, added to the accretion luminosity (\ref{Lacc}), are shown in Figures \ref{fig_fig3} and \ref{fig_fig4}.

They show the calculated evolution of the luminosity of the disk divided by the luminosity of the gainer over the relative existence time of the disk, normalized to unity, respectively for the short lived disk and liberally evolving medium mass binary $\beta$ Lyr and the longer lived disk and conservatively evolving low mass binary TT Hya. The vertical bar indicates the present disk with its present luminosity. It can be observed that in disks of Algol-type systems the accretion luminosity is of secondary importance as the luminosity originating from the gainer's radiation is two to three orders of magnitude larger. The present luminosity of the disk of $\beta$~Lyr (at 27~$\%$ of the disk life time) is Log~$\left(L_{disk}/L_{\odot}\right)$=4.21, about 71$\%$ the luminosity of the gainer. When 48$\%$ of the disk lifetime has been consumed the luminosity of the disk with a radius of 35~$R_{\odot}$ will exceed the luminosity of the gainer. The present luminosity of the disk of TT Hya (at 21$\%$ of the disk life time) is $L_{disk}/L_{\odot}=56$, about 90$\%$ the luminosity of the gainer. When 30~$\%$ of the disk lifetime has been consumed the luminosity of the large disk with a radius of 11.9~$R_{\odot}$ will exceed the luminosity of the gainer.

\begin{figure*}[!ht]
\centering
\includegraphics[width=10.6cm]{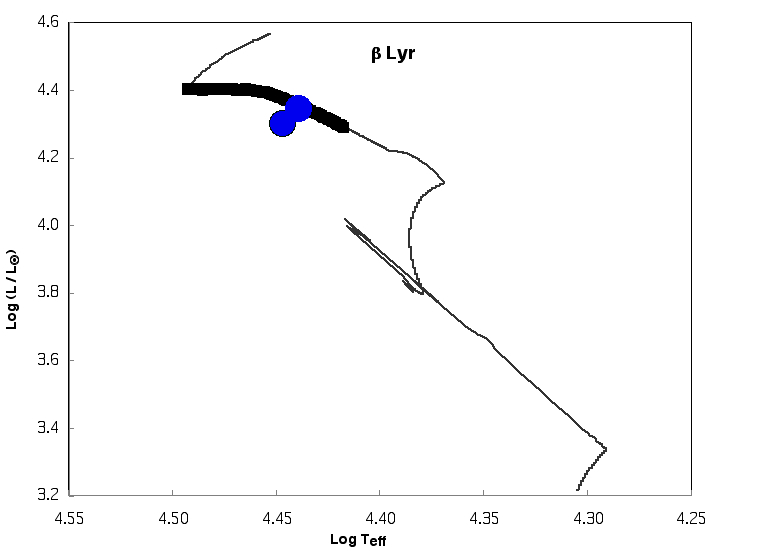}
\caption{Evolution of the gainer in $\beta$ Lyr through the HR-diagram. The short-lived disk is the thick part on the evolutionary path. The best fit model is the dot on the thick path whereas the observed position of the gainer is the dot to the lower left.}
\label{fig_fig5}
\end{figure*}

\begin{figure*}[!ht]
\centering
\includegraphics[width=9.6cm]{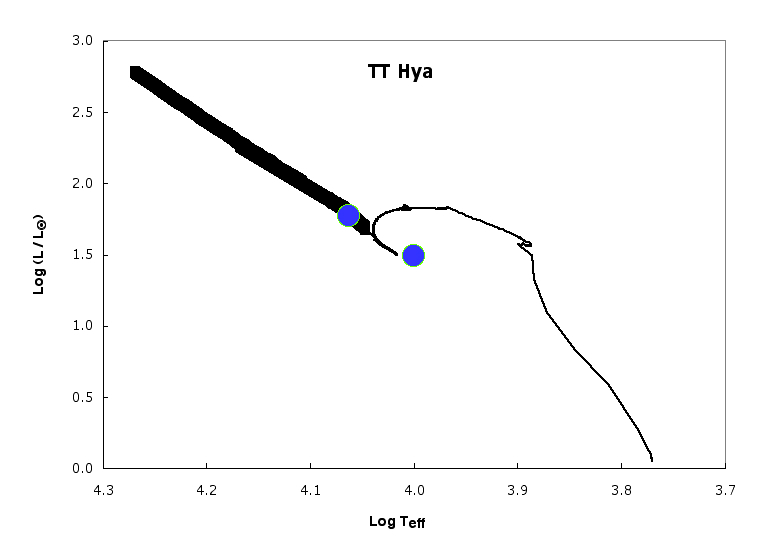}
\caption{Evolution of the gainer of TT Hya. The dot on the thick path is the best model, whereas the observed position of the gainer is to the right.}
\label{fig_fig6}
\end{figure*}

\vspace{0.2cm}

Figures \ref{fig_fig5} and \ref{fig_fig6} show the evolution of the gainer in the HRD for the $\beta$~Lyr and TT Hya. The disk phase is also indicated. $\beta$~Lyr has lived a Case A era of direct impact on the gainer. TT Hya started RLOF after exhaustion of hydrogen in the core of the donor. Both donors evolved through the main sequence and are now transferring mass during their hydrogen shell burning. As the orbit of $\beta$ Lyr is now rapidly widening a disk appears 2.63~$\times{10^{7}}$~y after ZAMS. 
This disk remains for 1.26~$\times{10^{5}}$~y, slightly more than 1$\%$ of the RLOF time. The system lives 9.551~$\times{10^{6}}$~y as an Algol. Previously published progenitors of  $\beta$~Lyr are from Ziolkowski (\cite{Ziolkowski}) (10 $M_{\odot}$ + 3.7 $M_{\odot}$, P=3.44d), De Greve $\&$ Linnell (\cite{DegreveLinnell}) (9 $M_{\odot}$ + 7.65 $M_{\odot}$, P=4d)~and Mennickent $\&$ Djurasevic (\cite{MennickentDjurasevic}) (12 $M_{\odot}$ + 7.2 $M_{\odot}$, P=2.5d). The present initial masses approximate those of De Greve $\&$ Linnell (\cite{DegreveLinnell}) calculated with external mass loss ($\Delta$M=0.45 $M_{\odot}$), but with a shorter initial period due to different values of external mass loss ($\Delta$M=1.5 $M_{\odot}$) and tidal interaction effects. The present model agrees  well with the progenitor proposed by Mennickent $\&$ Djuracevic (\cite{MennickentDjurasevic}), calculated with a somewhat larger amount of mass lost by the system.

In the case of TT Hya a disk appears 7.49~$\times{10^{8}}$~y after ZAMS. This disk remains for 1.60~$\times{10^{7}}$~y years, approximately 87$\%$ of the RLOF time. The system lives 1.76~$\times{10^{7}}$y as an Algol. 

\section{Conclusions}
\label{sec_Conclusions}

We extended our binary evolutionary code (Van Rensbergen et al. \cite{VanRensbergenetal1}, \cite{VanRensbergenetal2}) allowing for the action of viscous friction, tidal torque and meridional circulation. Matter leaving the system takes away the angular momentum of the gainer into space. Matter streams into interstellar space as a consequence of rapid rotation and the creation of high temperature accretion regions. The gauging of this event is still a matter of debate, depending on the well-known accretion luminosity and the poorly known size of the accretion region. Applying strictly the conservation of angular momentum we calculated the evolution of six  progenitors towards six well-know Algols with accretion disks. We followed the evolution of a progenitor into the presently defined physical data of the binary components and the accretion disk around the gainer. The temperature of the disk is determined by accretion and by the radiation from the gainer's surface. Fitting this rather large number of observed data allowed us to determine the masses and initial periods of the progenitors precisely. It has to be stated that different determinations of present mass, radii and effective temperatures of Algols would lead to different progenitors as found in this paper. Finally, we can summarize our findings as follows:

\vspace{0.1cm}
We obtained a more accurate determination of initial parameters of observed Algol systems with a disk, by adding the disk parameters in the comparison with the models.

\vspace{0.1cm}
Meridional circulation has to be included in the calculation of tidal interaction, at least for stars with radiative atmospheres.

\vspace{0.1cm}
The luminosity of accretion disks is determined by accretion upon and radiation from the gainer. However, the radiation luminosity is much larger than the accretion luminosity.

\vspace{0.1cm}
The luminosity of very large permanent disks exceeds the luminosity of the gainer at the end of RLOF.

\vspace{0.1cm}
None of the initial systems has a period longer than 2.7 days and our disk systems all have observed periods between 4.5 and 13 days. This reflects the fact that the systems evolved in a conservative or weakly liberal way.

\vspace{0.1cm}
Low values of the present mass ratio indicate that Algols that underwent RLOF during hydrogen core burning of the donor are at the end of their lives.

\vspace{0.1cm}
Algols that underwent their first RLOF during hydrogen shell burning of the donor are in the middle of their RLOF lives despite the low value of their mass ratio.

\begin{acknowledgements}

{We thank Monique \& Jean-Louis Tassoul for their advice concerning tidal interaction and Norbert Langer giving us a copy of the Ph.D. thesis of Stephan Wellstein and for his suggestion to use a spectroscopic HR-Diagram in order to evaluate the mass dependent values of the overshooting parameter. We also thank Mercedes Richards, who is no longer with us, for applying Doppler tomography for the determination of disk characteristics.}

 \end{acknowledgements}

\end{document}